\documentclass{svproc}
\usepackage{url}
\usepackage{cite}
\usepackage{amsmath,amssymb,amsfonts}
\usepackage{algorithmic}
\usepackage{graphicx}
\usepackage{textcomp}
\usepackage{xcolor}
\usepackage{comment}
\usepackage{float}

\begin{document}
\mainmatter
\title{Adapting atmospheric chemistry components for efficient GPU accelerators}
\titlerunning{GPU Chemistry}
\author{Christian Guzman Ruiz \inst{1} \and Matthew Dawson \inst{2} \and Mario C. Acosta \inst{1} \and Oriol Jorba \inst{1} \and Eduardo Cesar Galobardes \inst{3} \and Carlos Pérez García-Pando \inst{1} \and Kim Serradell \inst{1}
}
\authorrunning{Guzman Ruiz et al.} 
\institute{Barcelona SuperComputing Center, Barcelona, Spain\\
\email{christian.guzman@bsc.es},
\and National Center for Atmospheric Research (NCAR), Boulder, CO, USA \and Universitat Autonoma de Barcelona, Bellaterra, Spain}

\maketitle              

\begin{abstract}
Atmospheric models demand a lot of computational power and solving the chemical processes is one of its most computationally intensive components. This work shows how to improve the computational performance of the Multiscale Online Nonhydrostatic AtmospheRe CHemistry model (MONARCH), a chemical weather prediction system developed by the Barcelona Supercomputing Center. The model implements the new flexible external package Chemistry Across Multiple Phases (CAMP) for the solving of gas- and aerosol-phase chemical processes, that allows multiple chemical processes to be solved simultaneously as a single system. We introduce a novel strategy to simultaneously solve multiple instances of a chemical mechanism, represented in the model as grid-cells, obtaining a speedup up to 9$\times$ using thousands of cells. In addition, we present a GPU strategy for the most time-consuming function of CAMP. The GPU version achieves up to 1.2$\times$ speedup compared to CPU. Also, we optimize the memory access in the GPU to increase its speedup up to 1.7$\times$.
\end{abstract}

\section{Introduction}

Atmospheric models can be defined as a mathematical representation of the atmosphere's dynamical, physical, chemical, and radiative processes \cite{jacobson_fundamentals_2005}. They provide valuable information on the nature of real-world phenomena and systems, with many applications in science and engineering. However, they are often associated with large computational costs because of their complexity \cite{d_bennett_performance_2010}.

Due to the high computational cost, these models often divide their load into multiple parallel processes through domain decomposition \cite{tinto_optimizing_2017}. This technique divides a region into smaller regions, which will be called cells from now on. The model assigns collections of cells to independent threads to solve the many physical and chemical processes in the atmosphere in parallel. 

Atmospheric models use parallel programming interfaces like MPI, OpenMP, OpenACC, and CUDA to make this assignment. MPI is the most used tool to distribute work across independent supercomputer nodes. In addition, a model can use MPI to compute multiple cells in each computer node, and then use another parallel approach to further divide the load across individual CPU or GPU threads. Studies using the CUDA language have reported high speedups from parallelizing a demanding component of atmospheric models, the chemical kinetics module. For example, a study using the EMAC Earth system model developed a CUDA version of the Kinetic Preprocessor library (KPP) reporting a speedup of up to 20.4$\times$ against a single-thread execution \cite{alvanos_gpu-accelerated_2017}. Another research of simple chemical kinetics processes developed two different solver methods designed specifically for CUDA execution, the Runge-Kutta-Cash-Karp (RKCK) and Runge-Kutta-Chebysev (RKC) \cite{niemeyer_accelerating_2014}. This study achieved a speedup of up to 59$\times$ compared with a single-thread execution. The large difference in speedup between these studies highlights the importance of developing new methods specifically focused on parallel GPU execution and the impact of translating classic CPU-oriented methods to GPU. However, the GPU-specific methods are harder to adapt to atmospheric models and are often only tested for specific types of chemical equations. In contrast, CPU-based solvers, like KPP-GPU, are already prepared to run atmospheric models with the same chemical equations currently solved by purely CPU-based code.

The performance difference among these methods is primarily derived from their different parallelization methods. CPU-based solvers divide the load by domain decomposition, where each GPU thread solves an individual small system. The efficiency of this approach compared with CPU-only execution has been demonstrated multiple times \cite{haidar_guide_2017}. However, thousands of domain grid cells are required to speed up significantly. An alternative approach applied by the GPU-focused methods is to parallelize explicitly for chemistry equations. This approach allows for greater parallelization as each grid cell has multiple chemical reactions to solve. Also, these methods typically apply solving algorithms designed to execute more steps in parallel. These two characteristics of GPU-focused methods result in better performance than translation-based approaches.

This work presents results from solving simultaneously multiple cells in a single-thread execution. Also, we tested a GPU implementation following this strategy on the most time-consuming function of the chemical module. This approach combines the CPU-based and GPU-specific methods, as we still use a CPU-based solver, but use specific GPU techniques for one solver function to achieve a high degree of parallelization.

The proposed implementations are tested in the Chemistry Across Multiple Phases (CAMP) module. CAMP is developed to treat gas and aerosol chemical reactions in a single system, thus simplifying optimization and introduction of new multi-phase chemistry \cite{dawson_chemistry_2021}. It is integrated into the Multiscale Online Nonhydrostatic AtmospheRe CHemistry (MONARCH) model \cite{Badia2017}.

The remainder of this document is organized into sections according to these objectives. In section \ref{Background_ICICT}, we provide a brief description of MONARCH and CAMP, plus presenting the most time-consuming function of CAMP. In section \ref{Implementations_ICICT}, we present the GPU implementation of this function, an optimization to reduce GPU accesses, and the Multi-cells implementation for the whole CAMP module. In section \ref{Test_ICICT}, we define the hardware and software environment used. Section \ref{Results_ICICT} shows the result of the implementations presented. Finally, section \ref{Conclusions_ICICT} concludes the work and overviews possible future work.

\section{Background} \label{Background_ICICT}

MONARCH couples an online meteorological driver with gas and aerosol continuity equations to solve atmospheric chemistry processes in detail. The model is designed to account for feedbacks among gases, aerosol particles, and meteorology. This work focuses on its chemical components, which can consume up to 80\% of the model execution time. From the chemistry solvers available in MONARCH, we choose to work with the most and promising option, the framework CAMP.

CAMP is a novel framework permitting run-time configuration of chemical mechanisms for mixed gas- and aerosol-phase chemical systems (including gas- and aerosol-phase reactions and mass transfer), available at Github \cite{noauthor_camp_2021}. It also allows an abstract non-fixed representation of aerosols that can be configured at run-time, describing the life cycle of mineral dust, sea-salt, black carbon, organic matter (both primary and secondary), sulfate, nitrate, and ammonium aerosols. It computes a greater selection of types of chemical processes than the other MONARCH options. Thus, applying our implementations into CAMP affects more part of the chemistry time.

The chemical reactions in CAMP can include both integer parameters (e.g., array indices, stoichiometric coefficients, ionic charge, etc.) and floating-point parameters (e.g., conversion factors, rate parameters, etc.). The set of chemical species concentrations ($y$) is named the $state$ array, and the set of partial Derivatives of these species with respect to time ($f$) is named the $deriv$ array.

After the data is read, CAMP predicts future concentrations using the external ODE solver CVODE \cite{c._hindmarsh_sundials:_2004}. CVODE solves the time-dependent equation ( $y’ = f(t, y)$ ) using the CAMP-provided set of Derivatives ( $f(y)$ ) stored in the $deriv$ array. CVODE also uses a Jacobian matrix provided by CAMP. From the matrix structure options that CVODE offers, we choose the SPARSE structure \cite{skeel_construction_1986} to store the Jacobian, as this is a good choice for Jacobian structures with few non-zero elements, as is the case for many chemical mechanisms.

Either Derivative and Jacobian functions have very similar input and output, following the same structure. The only difference is the structure where we store the data (an array for the Derivative and a sparse matrix for the Jacobian) and some extra linear operations. So, we only need to analyze one of them since we can extrapolate the optimization ideas and techniques.

\begin{figure}
  \centering
  \includegraphics[width=0.5\linewidth]{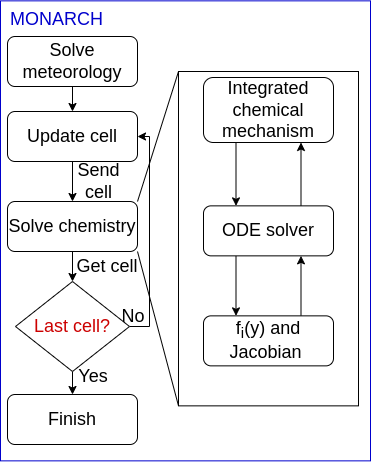}
      \caption{MONARCH overall flow diagram with CAMP as chemistry solver.}
  \label{Monarch_camp_2}
\end{figure}

Inside MONARCH, CAMP is required to solve chemistry multiple times. One time for each MONARCH time-step and cell. A cell represents a volume of the atmosphere, the collection of which composes a 3-dimensional grid that represents the whole atmosphere. The number of cells depends on the user-selected MONARCH configuration. MONARCH typically computes a large number of cells, over a large surface area with high precision. Each cell has its own $state$, which in terms of chemical processes is independent of other cell $state$ values. In figure \ref{Monarch_camp_2} we summarize the flow described in a diagram.

The CAMP functions executed during the solving takes a considerable execution time. We configured an experiment with a CB05 chemical mechanism to measure this impact. The experiment results show that CVODE occupies 70\% of the total execution time, and Derivative and Jacobian around 30\%. Despite being small functions compared to the whole ODE solver, the Derivative and Jacobian have a relevant impact on general performance, being the Derivative generally more time-expensive than Jacobian. So, in a similar way as selecting the chemistry component of MONARCH, we choose to work around the Derivative for analyzing our GPU implementation and search for a relevant reduction of the model execution time.

In general, chemistry models try to predict future concentrations of a set of chemical species by solving a set of ordinary differential equations that represent the reactions that compose a chemical mechanism. Reactions take the general form:

  \[c_1y_1 + \dots + c_my_m \leftrightarrow c_{m+1}y_{m+1} + \dots + c_ny_n,\]

\noindent where species $y_i$ is a participant in the reaction with stoichiometric coefficient $c_i$. The rate of change for each participating species $y_i$ with respect to reaction $j$ is given by

\[\left(\frac{dy_i}{dt}\right)_j =
  \begin{cases}
    \quad -c_ir_j(\mathbf{y},T,P,\dots) & \quad \text{for } i \le m \\
    \quad c_ir_j(\mathbf{y},T,P,\dots) & \quad \text{for } m < i \le n \\
  \end{cases},
\]

\noindent where the rate $r_j$ of reaction $j$ is an often complex function of the entire model state (including species concentrations $\mathbf{y}$, environmental conditions, such as temperature, $T$, and pressure, $P$, physical aerosol properties, such as surface area density and number concentration, etc.). The overall rate of change for each species $y_i$ at any given time is thus,

\[ f_i \equiv \frac{dy_i}{dt} = \sum_j \left(\frac{dy_i}{dt}\right)_j, \]

\noindent where $\mathbf{f}$ is referred to as the derivative of the system throughout this document.

Then, in the Derivative function we multiply the rate constants saved on the reaction parameters array with the corresponding concentrations on the $state$ array, filling the next concentration array ($deriv$). This operation is done for each reaction, adding all the results obtained from the reactions in the corresponding place of the $deriv$ array. So, we can say that each reaction adds a contribution to the $state$ concentrations, increasing or decreasing the value.

\section{Implementations} \label{Implementations_ICICT}

\begin{figure}
  \centering
  \includegraphics[width=\textwidth]{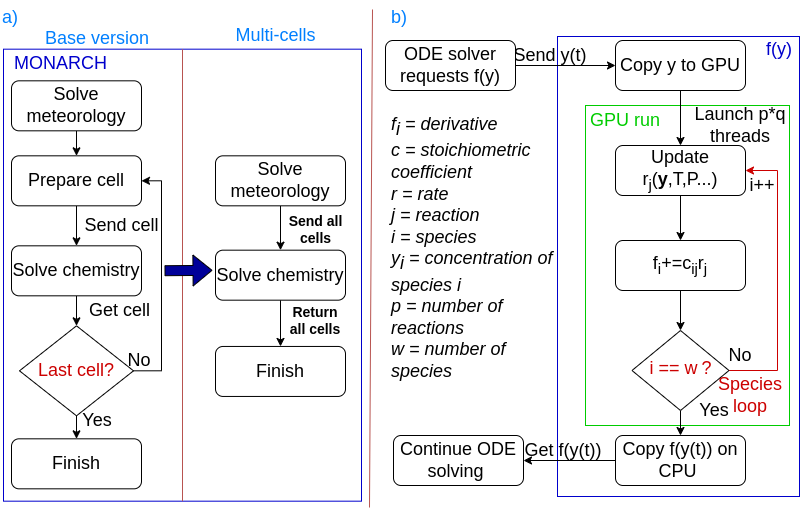}
      \caption{On the left (figure $a)$): comparison of original and Multi-cells overall workflows from the MONARCH point of view. On the right (figure $b)$): Derivative workflow diagram for GPU execution.}
  \label{Monarch_camp_gpu_deriv}
\end{figure}

The Multi-cells implementation groups the different input data from each cell into a single data structure to be computed. The MONARCH workflow described in figure \ref{Monarch_camp_2} is updated to figure \ref{Monarch_camp_gpu_deriv} part $a)$. The cells loop disappear inside the solving internal functions, avoiding the process of updating the input data from cells and re-initializing the ODE solver. As an example, the Derivative equation is updated as following:

\[ f_i \equiv \frac{dy_{ik}}{dt} = \sum_j \left(\frac{dy_{ik}}{dt}\right)_j \]

where $y_ik$ refers to the species $y_i$ from cell $k$.

Our GPU strategy is the parallelization of each reaction data packet. In figure \ref{Monarch_camp_gpu_deriv} part $b)$, we can see the resultant GPU-based Derivative flow diagram.

We compute the sum of contributions to f by using the CUDA operation $atomicAdd$. This function avoids a possible thread overlapping on updating the same variable. This interference can be produced by reactions with common species between them.

Reaction data is allocated on global memory at the initialization of the program. To send and receive from the GPU the rest of the data ($state$ array), we check first the size of this array. If it contains few data variables, then $state$ is passed as a function parameter, taking advantage of the constant memory. Otherwise, the data is copied through a direct transfer to the global memory.

The number of GPU threads initialized is equal to the number of reactions. Another relevant GPU parameter, the number of blocks per threads, is configured to the maximum available for the GPU used (1024 threads/block). Lower configurations of threads/blocks don’t show a performance improvement for our tests. Due to the possibility of using a GPU with less capacity in the future, we add a run-time checking of GPU hardware specifications to ensure the correct execution of the program regardless of the GPU used (for example, avoid demanding more threads than the GPU limit).

\begin{figure}
  \centering
  \includegraphics[width=0.7\textwidth]{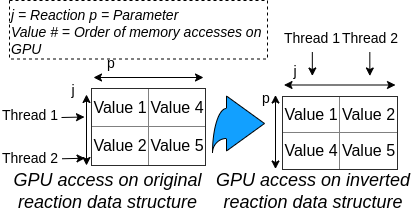}
      \caption{Data structure inversion for GPU Derivative. “Value” numbers represent the GPU memory arrangement and access order, “j” the number of reactions and “p” the number of parameters.}
  \label{Reverse_matrix}
\end{figure}

In the still CPU-based implementation all the reaction data packets are initially stored consecutively in memory. Then, the parallelization by reactions results in each thread accessing no-consecutive values of the reaction data structure. We reordered this structure to follow a sequential reading of the data in the GPU. The first reaction parameters accessed are stored consecutively in the reaction data structure, and so on. Figure \ref{Reverse_matrix} illustrates the changes in the data packet structure, simulating the structure as a matrix where initially the rows are the data packets and columns the parameter values.

\section{Test environment} \label{Test_ICICT}
During the work, all the tests and executions were performed in the CTE-POWER cluster provided by the Barcelona Supercomputing Center (BSC) \cite{noauthor_support_nodate}. We used the compilers GCC version 6.4.0. and NVCC version 9.1, an IBM Power9 8335-GTH @ 2.4GHz and a GPU NVIDIA V100 (Volta) with 16GB HBM2.

We work around a basic chemical mechanism of 3 species, where species $A$ generates $B$ and $C$ through 2 Arrhenius reactions. $A$ is initialized at 1.0, while $B$ and $C$ are set to zero. Each cell has a small offset of 0.1 on the initial concentrations to generate different results. For example, at the first concentration value, we sum a 0.1 offset value, at the second 0.2, and so on till Multi-cells species. The rest of the variables, like temperature, pressure, or reaction data parameters, are initialized to the same values for all the cells.

\section{Results} \label{Results_ICICT}

\begin{figure}
  \centering
  \includegraphics[width=\textwidth]{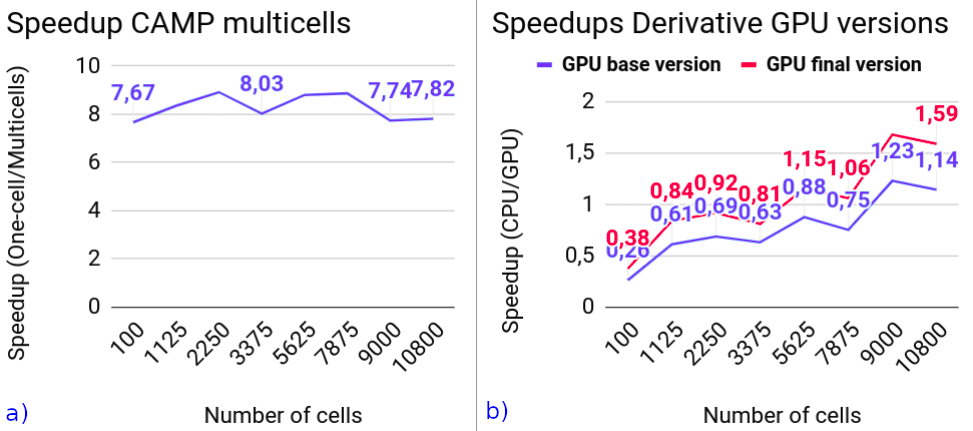}
      \caption{On the left (figure $a)$): CAMP speedup using Multi-cells optimization in front of the original One-cell version. On the right (figure $b)$): Speedup of base and final single-GPU versions compared to single-thread CPU versions. Final version applies the optimization on GPU memory access into the base version.}
  \label{speedup_deriv_and_camp2}
\end{figure}

In figure \ref{speedup_deriv_and_camp2} part $a)$, we can see how Multi-cells speedups CAMP a factor of 8$\times$ for multiple numbers of cells. Most of this speed-up is produced by the reduction of solving iterations. In the One-cell case the number of iterations scales linearly with the number of cells factor, while in the Multi-cells case the number of iterations is independent from the number of cells computed, keeping almost the same number of iterations for the number of cells. For example, One-cell takes around 6e$^{6}$ iterations to solve 10,000 cells (an average of 600 iterations per cell), while Multi-cells takes around 700 to solve all cells, with independence of the number of cells.

The GPU implementation speeds up the Derivative function for a high number of cells. In figure \ref{speedup_deriv_and_camp2} part $b$, we can see how for 10,000 cells the GPU version achieves 1.2$\times$ speedup. On the other hand, a lower number of cells slowdowns the function, shown as a speedup below 1$\times$ for less than 10,000 cells. We can also see that the optimization on memory access improves the overall speedup by a factor of 1.3$\times$ approximately for all numbers of cells.

We also compare the final GPU version against a CPU case parallelized with MPI, emulating the parallelization used in MONARCH. The number of MPI processes are configured to follow the proportion of GPUs used for GPU available. So, we use 40 MPI threads from the 160 available, like the GPU experiments presented use 1 GPU from the 4 available. We obtain that the GPU execution is 3x times slower than the MPI, but only because the time of data movements between CPU and GPU is taking near ~90\% of the GPU execution time. The GPU computation time is 3.5$\times$ times faster than the MPI time (0.04s in GPU and 0.14s for MPI). This data movement is produced by updating the species concentrations on each call to Derivative. We can conclude that the GPU Derivative function has a small computation load for data movement produced (reaction data, concentration values, etc.)

\section{Conclusions} \label{Conclusions_ICICT}

In this paper we focused on improving the performance of CAMP for an execution in an atmospheric model environment like MONARCH. MONARCH simulations perform one CAMP simulation for each grid cell of the geographic simulation region for each MPI thread. These cells have no inter-dependencies during the chemistry solving; thus they have potential to be parallelized by the GPU. However, the classical MONARCH implementation calls the CAMP solving process for each grid cell. In each cell iteration, the CAMP solving library (CVODE) needs to reinitialize its internal solving variables. Furthermore, to implement a GPU implementation over the cells it would be necessary to translate the complete solving code to GPU format, which can be an exhaustive work. The first implementation presented in this study aims to solve these issues. This strategy is relatively novel in the atmospheric community and can be used as an example to speed-up the model. We refer in the paper to this implementation using the name of Multi-cells.

The Multi-cells strategy groups the data for each cell into a single structure to be solved. The cells loop from MONARCH is moved into the internal solving functions of CAMP. The results show a considerable reduction of the calls to the Derivative function. The solving module uses approximately the same number of iterations to solve all the cells than to solve a single cell. With respect to the improvement in execution time, the Multi-cells implementation achieves near 8$\times$ speedup for all the cells tested, up to 9$\times$ speedup.

Next, we developed a CUDA version of the Derivative function by parallelizing its reaction loop among GPU threads. The new version obtains near 1.2$\times$ speedup for 10,000 cells approximately. For a lower number of cells, the CPU version has better performance than GPU. The third implementation reorder the reaction data structure to improve its access in the GPU Derivative version, increasing the GPU speedup by a factor of 1.3$\times$ for all the cells tested. 

Finally, we inspect the time execution consumed on moving data between GPU and CPU. For 10,800 cells, this time on data movement takes ~90\% of the total time execution. Comparing the results with a 40 MPI process execution, the computation time for the GPU version is 3.5$\times$ faster. Thus, future work will focus on reducing GPU data movement by translating more CPU functions to the GPU, for example the Jacobian or functions from the ODE solving and overlapping some CPU and GPU work. This should increase the computation performed on the GPUs and reduce data movement by transferring data only at the start and the end of the solving, reducing data movement during solver iterations. This can be done by parallelizing the next solver functions executed after or before the Derivative calculation until all the solver would be executed in GPU. We also expect to explore load balancing the CPU and GPU using overlapping and asynchronous communication, since currently, the CPU is not performing any work during GPU execution. Lastly, we expect to evaluate the GPU-based chemistry solving in MONARCH, checking the impact for a variety of atmospheric experiments with an MPI implementation alongside the GPU–CUDA chemistry.

\section*{Acknowledgment}
This work was partially supported by funding from the Ministerio de Ciencia, Innovación y Universidades as part of the BROWNING project (RTI2018-099894-BI00), the CAROL project (MCIN AEI/10.13039/501100011033 under contract PID2020-113614RB-C21), the Generalitat de Catalunya GenCat-DIUiE (GRR) (2017-SGR-313) and the AXA Research Fund through the AXA Chair on Sand and Dust Storms established at BSC. This work has also received funding from "Future of Computing Center, a Barcelona Supercomputing Center and IBM initiative (2020)". Matthew Dawson has received funding from the European Union’s Horizon 2020 research and innovation program under Marie Skłodowska-Curie grant agreement no. 747048. This paper expresses the opinions of the authors and not necessarily those of the funding commissions. BSC co-authors acknowledge the computer resources at CTE-POWER, the technical support provided by the Barcelona Supercomputing Center, and the support from Partnership for Advanced Computing in Europe (PRACE) and Red Española de Supercomputacion (RES).

\bibliographystyle{splncs_srt}
\bibliography{main.bib}

\end{document}